\documentclass[12pt]{article}

\usepackage{sbc-template-br}
\usepackage[square]{natbib}

\usepackage{graphicx}
\usepackage{subfigure}
\usepackage{url}

\usepackage[brazil]{babel}   
\usepackage[utf8]{inputenc}  

\usepackage{paralist}
\usepackage{url}

 \usepackage[empty]{fullpage}
\usepackage[usenames]{color}

\usepackage{listings}
\title{Expressando Atributos Não-Funcionais em Workflows Científicos}

\author{
Vívian Medeiros\and
Antônio Tadeu Azevedo Gomes
}

\address{
Laboratório Nacional de Computação Científica (LNCC) -- Petrópolis, RJ -- Brasil\\
\vspace{-0.3in}
\email{\{vivian, atagomes\}@lncc.br}  
}

\begin{document} 

\maketitle

\begin{abstract}
In this paper we present 
OSC, a scientific workflow specification language based on software architecture principles.
In contrast with other approaches, OSC employs connectors as first-class constructs.
In this way, we leverage reusability and compositionality in the workflow modeling process, specially in the configuration of mechanisms that manage non-functional attributes.
\end{abstract}
     
\begin{resumo} 
Este artigo apresenta 
OSC, uma linguagem de especificação de workflows científicos baseada em princípios de arquitetura de software.
Em contraposição a outras abordagens, OSC emprega conectores como construções de primeira classe. 
Desse modo, propicia-se uma maior capacidade de reuso e composicionalidade na modelagem de workflows, particularmente nas configurações dos mecanismos que lidam com atributos não-funcionais.
\end{resumo}

\section{Introdução}

Trabalhos recentes sobre sistemas gerenciadores de workflows científicos (SGWfCs) têm demonstrado que a comunidade científica vem se preocupando em adicionar suporte a atributos não-funcionais
a esses sistemas~\citep{ludascher2006,mouallem2010,gadelhajr2011}. 
No entanto, tais atributos comumente não podem ser especificados nos modelos dos workflows, pelo fato das linguagens de especificação de workflows existentes possuírem expressividade em geral limitada para esse fim.
Essa característica torna a modelagem dos workflows mais simples, porém oferece menor flexibilidade na configuração dos mecanismos associados a esses atributos. 
Quando o sistema oferece suporte à configuração desses atributos, a mesma é concentrada na especificação das tarefas (componentes computacionais), ou é associada ao workflow como um todo.
Essa característica dificulta ou impossibilita a configuração desses mecanismos nas comunicações e coordenações empregadas entre tarefas.

Em busca do aprimoramento dessa expressividade, este trabalho apresenta a linguagem 
OSC, uma evolução do trabalho preliminar apresentado como resumo estendido em~\citep{medeiros2011}. 
OSC é definida 
sobre a linguagem de descrição arquitetural Acme~\citep{garlan1997}.
Em contraposição a outras abordagens,
OSC emprega \textbf{conectores} como construções de primeira classe para a 
modelagem tanto de tipos quanto de instâncias de intera-ções entre 
tarefas quanto de regras que governam essas interações.
Com essa abordagem, OSC propicia uma maior capacidade de reuso, composicionalidade e configurabilidade na modelagem de workflows, beneficiando particularmente 
o tratamento de atributos não-funcionais.

O restante deste artigo está estruturado como se segue. 
A Seção \ref{sec:atr_nf} apresenta os atributos não-funcionais tratados neste trabalho. 
A Seção~\ref{sec:osc} apresenta os elementos de modelagem de OSC.
A Seção~\ref{sec:ex} apresenta um exemplo de uso de OSC,
que é comparada a trabalhos relacionados na Seção~\ref{sec:trab}. 
Por fim, na Seção~\ref{sec:conc} são apresentadas as conclusões.

\section{Atributos não-funcionais em workflows científicos}
\label{sec:atr_nf}

O levantamento dos atributos não-funcionais tratados neste trabalho foi realizado a partir da análise de workflows científicos existentes (como o OrthoMCL~\citep{fischer2011} e o ProFrager, este último apresentado na Seção \ref{sec:ex}) e de alguns dos SGWfCs (vide Seção \ref{sec:trab}) mais populares na literatura dentre aqueles que permitem a composição e configuração destes atributos em uma linguagem de modelagem (seja ela gráfica ou textual). 
Neste trabalho são tratados como atributos não-funcionais: 
\begin{inparaenum}[(i\upshape)]
\item os atributos de qualidade relacionados a confiabilidade e rastreabilidade,
\item o paralelismo de tarefas, e
\item o paralelismo de dados.
\end{inparaenum}
Nesse levantamento, o escalonamento de tarefas também se mostra um atributo não-funcional importante.
Porém, como o mesmo deve ser tratado fim-a-fim em qualquer workflow e sua configuração depende de informações contidas na descrição das tarefas e conectores, escolheu-se tratá-lo diretamente no SGWfC que executa workflows OSC.
A implementação de um SGWfC para OSC existe e está disponível (vide Seção~\ref{sec:ex}), mas seu detalhamento foge ao escopo deste trabalho.

Falhas podem ocorrer em diversas partes de um workflow, podendo ser falhas tanto nas tarefas quanto em suas interações, e por diversos motivos, como falhas nas transferências de dados ou falta de bibliotecas necessárias à execução de tarefas. Essa característica ressalta a importância da adoção de mecanismos de tolerância a falhas em SGWfCs de forma a adicionar confiabilidade às execuções.
Já os mecanismos de rastreamento de proveniência de dados são utilizados por SGWfCs para uma melhor gerência dos metadados que podem ser gerados em cada execução de um workflow. 
Workflows podem gerar uma quantidade significativa de metadados, 
o que tem estimulado a comunidade científica a buscar soluções que facilitem essa gerência em SGWfCs \citep{gadelhajr2011}. 

Ambientes para execução paralela de software têm sido crescentemente associados a SGWfCs.
Dois tipos principais de paralelismo de tarefas são em geral considerados: 
memória compartilhada e 
memória distribuída. 
Apesar de aceleradores (como GPUs) serem uma tendência,
optou-se por não abordá-los inicialmente neste trabalho, pois os exemplos de workflows estudados não apresentaram nenhuma tarefa que dependesse deste tipo de paralelismo.

Workflows podem ser usados para o processamento de grandes massas de dados. 
Os esquemas de varredura de parâmetros e MapReduce são interessantes para esse tipo de processamento quando os dados podem ser divididos para o processamento (em geral, paralelo) de conjuntos menores de dados. 
A varredura de parâmetros consiste em invocações repetidas de uma tarefa utilizando diferentes dados de entrada para cada invocação, 
podendo portanto ser usada também 
em simulações computacionais 
baseadas em métodos como o de Monte Carlo.
Já no MapReduce \citep{dean2008} uma função \textit{map} processa um par \{chave,valor\} e gera um conjunto intermediário de pares \{chave,valor\}. 
Uma função \textit{reduce} processa todos os pares gerados pela função \textit{map} com uma mesma chave. 
Para gerar os pares de entrada da função \textit{reduce}, após a função \textit{map} é executada uma fase intermediária de ordenação das chaves e fusão dos valores regidos pela mesma chave. 

\section{OSC: \textit{Open Scientific Connectors}}
\label{sec:osc} 

OSC é definida sobre 
a linguagem Acme~\citep{garlan1997}. 
Em Acme é possível descrever estilos arquiteturais 
que permitem o reuso de elementos de modelagem em diferentes arquiteturas de software, bem como a definição de regras de composição desses elementos. 
Um estilo foi definido em Acme para a descrição dos elementos -- \textbf{tarefas}, \textbf{conectores}, \textbf{portas}, e \textbf{papéis} -- e regras de modelagem de workflows em OSC.
Em OSC, tarefas só se comunicam por meio de conectores. 
Tarefas e conectores possuem interfaces denominadas, respectivamente, de portas e papéis. 
O modelo de um workflow em OSC envolve a ligação de portas de entrada/saída de tarefas a papéis de origem/destino de conectores. 
Ligações entre portas e papéis podem representar dependências  de controle ou de dados entre tarefas.

OSC considera a existência de dois tipos de usuários no processo de especificação de workflows científicos: \emph{cientistas} e \emph{projetistas}. 
O usuário \emph{cientista} descreve workflows em termos de relações entre instâncias de tipos pré-definidos de tarefas e de conectores (desenvolvimento \emph{com} reuso).
O usuário \emph{projetista} pode estender OSC definindo novos tipos de tarefas e conectores com base nos tipos pré-definidos pela linguagem (desenvolvimento \emph{para} reuso). 

OSC predefine um conjunto de tipos básicos para tarefas, portas, conectores e papéis.
Esses tipos básicos associam o elemento de modelagem abstrato no workflow com sua imple-mentação concreta. 
Por exemplo, uma tarefa pode ser um executável ou um ``fluxo'' (um workflow encapsulado como uma tarefa), enquanto um conector pode ser um \textit{pipe} de caracteres ou o transporte de um arquivo.
Associado a esses tipos básicos são predefinidos também tipos específicos para representar a configuração 
de atributos não-funcionais. 

\begin{figure}[h!]
\begin{center}
{\subfigure[Tipos básicos de Tarefa]{\label{subfig:bas_tar}\includegraphics[width=0.29\textwidth]{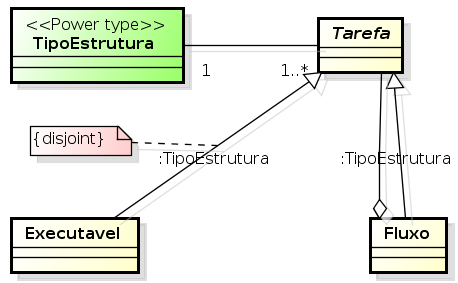}}}
{\subfigure[Extensão do tipo Executavel]{\label{subfig:exec_est}\includegraphics[width=0.29\textwidth]{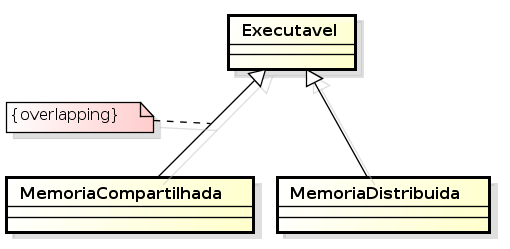}}}
{\subfigure[Extensão do tipo Fluxo]{\label{subfig:fluxo_est}\includegraphics[width=0.29\textwidth]{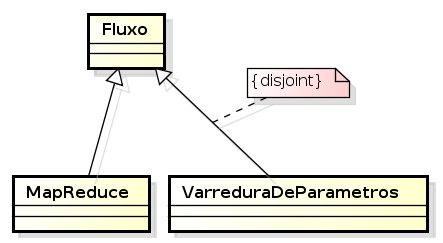}}}
\caption{Tipos básicos e específicos de tarefas em OSC}
\label{fig:tarefas}
\end{center}
\end{figure}

A Figura \ref{fig:tarefas} apresenta diagramas UML que representam os tipos básicos de tarefa e alguns de seus tipos específicos.
Os diagramas usam o formato proposto por \cite{Hnatkowska2005}.
Optou-se por usar generalizações, restrições e \textit{powertypes} da UML 2.0 para retratar esses tipos neste artigo, ao invés das especificações Acme correspondentes, devido ao espaço disponível.
Contudo, para ilustrar o uso de Acme 
alguns trechos dessas especificações são apresentados nas subseções que se seguem. 

%
%

A Figura \ref{subfig:bas_tar} define o \textit{powertype} TipoEstrutura para englobar os tipos básicos de tarefas, os quais não podem ser combinados por se tratarem de tipos disjuntos. 
A Figura \ref{subfig:exec_est} apresenta os tipos OSC referentes ao atributo não-funcional de paralelismo de tarefas.
A Figura \ref{subfig:fluxo_est} apresenta os tipos OSC referentes ao atributo não-funcional de paralelismo de dados.
Com exceção do tipo VarreduraDeParametros, todos os outros atributos não-funcionais representados na Figura \ref{fig:tarefas} são exclusivos para o tipo Tarefa. 
Portas também possuem tipos para a varredura de parâmetros, de forma a permitir a configuração de bifurcações e junções.

\begin{figure}[h!]
\center
\includegraphics[width=0.7\textwidth]{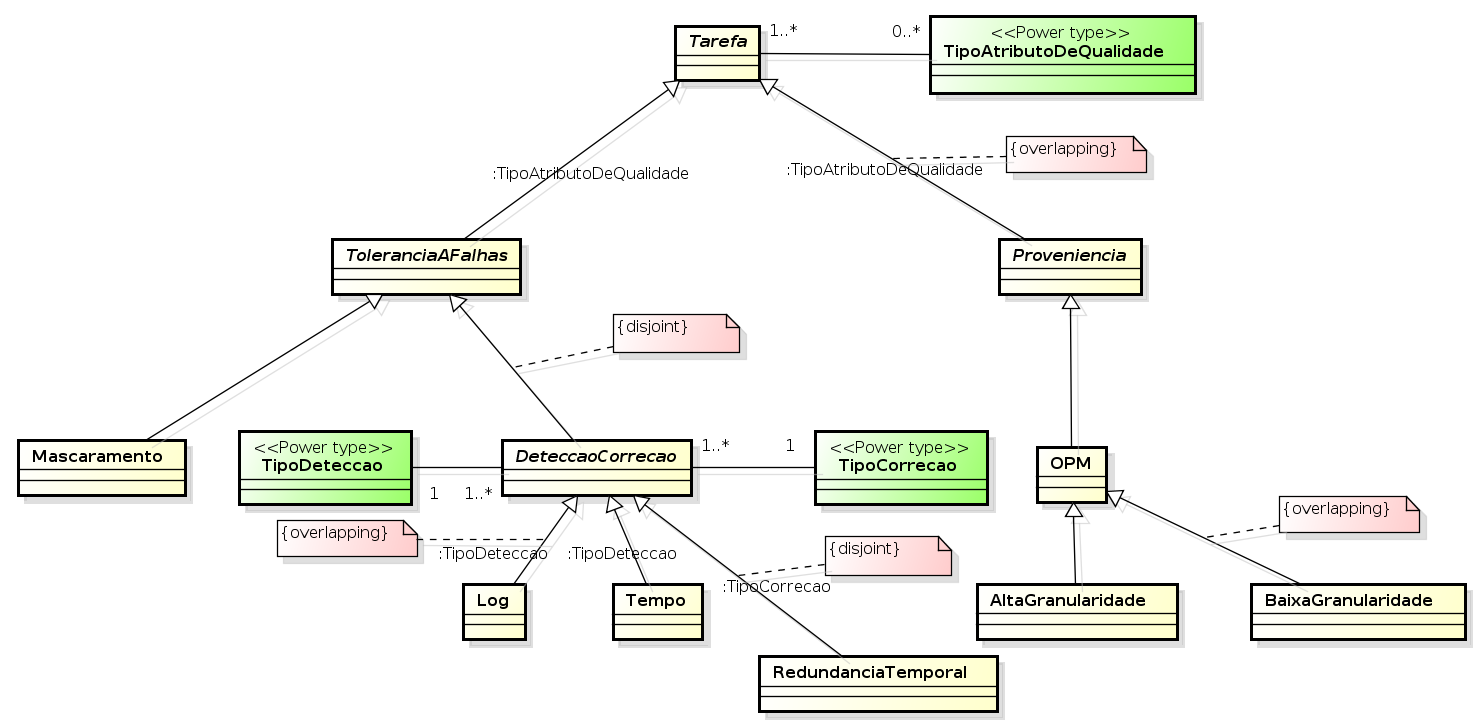}
\caption{Diagrama de tipos de tipos atributos de qualidade para tarefas}
\label{fig:atr_tarefas}
\end{figure}

\begin{figure}[h!]
\center
\includegraphics[width=0.5\textwidth]{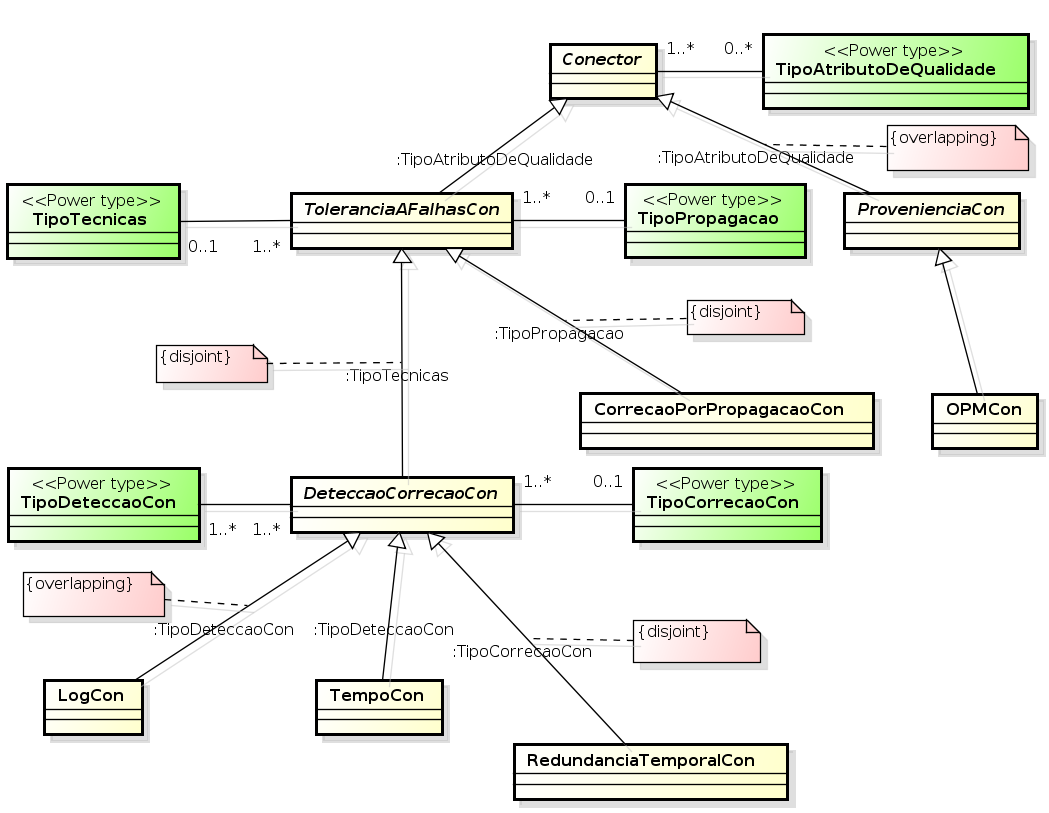}
\caption{Diagrama de tipos atributos de qualidade para conectores}
\label{fig:atr_conectores}
\end{figure}

As Figuras \ref{fig:atr_tarefas} e \ref{fig:atr_conectores} mostram diagramas UML que representam os tipos de atributos de qualidade para tarefas e conectores. 
Em OSC os atributos de qualidade são classificados pelo \textit{powertype} TipoAtributoDeQualidade. 
Todos os elementos OSC estão associados a esse \textit{powertype}, porém nem todos os atributos de qualidade são tratados em todos os elementos e o tratamento é distinto em cada elemento. 
Os parágrafos a seguir 
apresentam mais detalhes sobre como os atributos não-funcionais são tratados em OSC.

\vspace{-1in}
\paragraph{Paralelismo de tarefas.}
Os tipos MemoriaCompartilhada e MemoriaDistribuida (vide Figuras \ref{subfig:exec_est} e \ref{osc_tipomemoria})
permitem que tarefas paralelas 
sejam adicionadas ao fluxo. 
Estas tarefas podem executar em sistemas computacionais que utilizam diferentes gerenciadores locais de recursos. 
Nesse sentido, optou-se neste trabalho por uma abordagem minimalista, na qual 
o tipo MemoriaCompatilhada permite somente a configuração do número de \textit{threads} que serão disparadas e o tipo MemoriaDistribuida permite somente a configuração do número de nós no sistema computacional que será usado para a execução e o número de processos que serão disparados em cada nó. 
Através da combinação destes tipos de alocação, OSC dá suporte a configuração de executáveis que implementem paralelismo híbrido (p. ex., combinando \textit{pthreads} e MPI).

\vspace{-0.2in}
\paragraph{Paralelismo de dados.}
Os tipos VarreduraDeParametros e MapReduce herdam do tipo básico de tarefa Fluxo. 
O tipo VarreduraDeParametros, descrito na Figura~\ref{osc_tipovarredura},  pode ser configurado para executar seu mecanismo de forma sequencial ou pela criação de instâncias paralelas de suas tarefas internas. 
Para que uma varredura de parâmetros possa ocorrer ao menos uma porta de entrada deste tipo de fluxo deve ser do tipo Bifurcacao. 
Esse tipo de porta deve ser associado ou a um conjunto de dados (diretório de arquivos ou lista de valores) ou a um número de instâncias do fluxo 
a serem executadas. 
O conjunto de dados ou o número de instâncias de cada porta de bifurcação define se o SGWfC deve realizar a combinação de dados e/ou a repetição dos experimentos, permitindo a criação da quantidade de instâncias das tarefas do fluxo que deverão ser executadas.
Todas as portas de saída de dados de um fluxo do tipo VarreduraDeParametros devem ser do tipo Juncao, o qual é responsável pela junção dos dados de saída gerados após o término de todas as instâncias da varredura de parâmetros. 
Existem três formatos possíveis para junção:
\begin{inparaenum}[(i\upshape)]
 \item \textit{include}, que adiciona arquivos a um diretório;
 \item \textit{merge}, que adiciona o conteúdo de um diretório em outro diretório; e
 \item \textit{concat}, que concatena arquivos em outro arquivo. 
\end{inparaenum}
No que se refere ao MapReduce, diversos sistemas (p.ex.\ Hadoop e BashReduce \citep{silva2011}) provêem implementações distintas para esse modelo, 
o que dificulta a generalização deste tipo. 
Por isso, em OSC é adotada uma representação para este tipo em que um fluxo possui um conjunto de binários que se responsabilizam pela execução do MapReduce. 
Apesar deste tipo não estar relacionado a um mecanismo específico, ele propicia uma melhor legibilidade do modelo do workflow.

\begin{figure}[h!]
\begin{center}
{\subfigure[Tipos para paralelismo de tarefa]{\label{osc_tipomemoria}
\begin{lstlisting}
Component Type MemoriaCompartilhada extends Executavel with {
        Property num_threads : int;
	//... regras do tipo
}
Component Type MemoriaDistribuida extends Executavel with {
        Property num_nos : int;
        Property num_procs_por_no : int;
	//... regras do tipo
}
\end{lstlisting}}}
\vline
{\subfigure[Tipos para paralelismo de dados]{\label{osc_tipovarredura}
\begin{lstlisting}
Property Type JuncaoT = Enum {include,merge,concat};
Component Type VarreduraDeParametros extends Fluxo with {
	Property exec_seq : boolean;
	//... regras arquiteturais
	//... regras do tipo	
}
Port Type Bifurcacao extends PortaVarreduraDeParametros with {
    Property num_copias : int;
    //... regras arquiteturais
    //... regras do tipo
}
Port Type Juncao extends PortaVarreduraDeParametros with {
    Property tipo_juncao : JuncaoT;
    //... regras arquiteturais
    //... regras do tipo
}
\end{lstlisting}}}
\caption{Tipos específicos de tarefas em OSC}
\label{fig:paralelismo}
\end{center}
\end{figure}

\vspace{-0.2in}
\paragraph{Tolerância a falhas.}
OSC permite o tratamento de falhas tanto em tarefas quanto em conectores, como pode ser visto na Listagem~\ref{osc_tipotoleranciafalhas}. 
Neste trabalho, as técnicas mascaramento e detecção/correção foram usadas para prover tolerância a falhas. 
O mascaramento representa a redundância de hardware e é usado somente por tarefas. 
Nele diversas cópias da tarefa são executadas simultaneamente e, ao final das execuções, o conjunto de resultados é analisado por algoritmos de votação de forma a gerar o resultado final.
A detecção/correção consiste em duas etapas, onde a primeira detecta a falha e gera um sinal para que a segunda possa tentar corrigir o problema. 
Uma instância de tarefa ou conector que use essa técnica obrigatoriamente precisa combinar um tipo de detecção a um tipo de correção.
As técnicas oferecidas atualmente por OSC para detecção de falhas são:
\begin{inparaenum}[(i\upshape)]
 \item análise de log, que gera um sinal de erro quando uma falha é detectada nos logs da tarefa; e 
 \item monitoramento do tempo, onde a execução (em tarefas) ou a transferência de dados (em conectores) é monitorada e um sinal de erro é criado quando o tempo limite para a operação é excedido. 
\end{inparaenum}
Falhas de tarefas que sejam sinalizadas através de portas de saída podem ser tratadas em conectores pela técnica de correção por propagação, a qual é usada quando a falha da tarefa não é prejudicial à execução do workflow como um todo.
Nesse caso, o conector recebe em seu papel de origem o sinal de falha da porta de saída da tarefa e garante a continuidade do fluxo descartando os dados de saída da tarefa que apresentou erro. 
Outra técnica de correção oferecida na linguagem OSC é a redundância temporal, na qual a execução da tarefa ou a transferência de dados pelo conector que apresenta falha é abortada, sendo realizadas no máximo $N$ novas tentativas de execução/transferência (sendo $N$ configurável na descrição do workflow).

\vspace{-0.23in}
\paragraph{Proveniência de dados.}
A descrição de tipos de configuração de proveniência no OSC (apresentados na Figura~\ref{osc_tipoproveniencia}) é baseada nas definições do formato \textit{Open Provenance Model} (OPM) \citep{moreau2011}.
Todos os elementos de modelagem em OSC 
podem ser combinados ao tipo OPM. 
A propriedade \textit{versao} está presente em todos esses elementos e permite a geração de diversas descrições de uma mesma execução em um único grafo OPM. \footnote{Detalhes do versionamento de grafos OPM podem ser encontrados em \cite{moreau2011}.}
Os tipos AltaGranularidade e BaixaGranularidade se aplicam somente a fluxos e permitem a definição da granularidade de proveniência. 
Um fluxo do tipo AltaGranularidade permite que sua representação interna tenha proveniência armazenada de forma a considerar todos os elementos do tipo OPM que encapsula. 
Um fluxo do tipo BaixaGranularidade considera o fluxo como uma única tarefa. 
Esses tipos podem ser combinados entre si na criação de diversas versões do grafo OPM.

\begin{figure}[h!]
\begin{center}
{\subfigure[Tipos para tolerância a falhas]{\label{osc_tipotoleranciafalhas}
\begin{lstlisting}
Component Type ToleranciaAFalhas 
		extends Tarefa with {}
Component Type Mascaramento 
		extends ToleranciaAFalhas with {
        Property num_copias : int;
        Property algoritmo : string;
}
Component Type DeteccaoCorrecao 
		extends ToleranciaAFalhas with {}
Component Type Log 
		extends DeteccaoCorrecao with {}
Component Type Tempo 
		extends DeteccaoCorrecao with {
	Property tempo_max : int;
}
Component Type RedundanciaTemporal 
		extends DeteccaoCorrecao with {
	Property num_tentativas : int;
	Property ignorar : boolean;
}
Connector Type ToleranciaAFalhasCon 
		extends Conector with {}
Connector Type DeteccaoCorrecaoCon 
		extends ToleranciaAFalhasCon with {}
Connector Type LogCon 
		extends DeteccaoCorrecaoCon with {}
Connector Type TempoCon 
		extends DeteccaoCorrecaoCon with {
        Property tempo_max : int;
}
Connector Type CorrecaoPorPropagacaoCon 
		extends ToleranciaAFalhasCon with {}
Connector Type RedundanciaTemporalCon 
		extends DeteccaoCorrecaoCon with {
        Property num_tentativas : int;
}
\end{lstlisting}}}
\vline
{\subfigure[Tipos para rastreamento de proveniência]{\label{osc_tipoproveniencia}
\begin{lstlisting}
Property Type VersaoT = Set {string};
Property Type ArmazenamentoOPMT = Enum {xml,rdf};
Component Type OPM extends Proveniencia with {
	Property versao : VersaoT;
	Property anotacao : string;
}
Component Type AltaGranularidade extends OPM with {
	Property serializar_grafo_por_processo : boolean;
	Property armazenamento : ArmazenamentoOPMT;
}
Component Type BaixaGranularidade extends OPM with {
	Property armazenamento : ArmazenamentoOPMT;
}
Port Type PortaOPM extends PortaProveniencia with {
        //... propriedades
}
Role Type PapelOPM extends PapelProveniencia with {
        //... propriedades
}
Connector Type OPMCon extends ProvenienciaCon with {
	//... propriedades
}
\end{lstlisting}}}
\caption{Tipos de atributos de qualidade em OSC}
\label{fig:attr_qualidade}
\end{center}
\end{figure}

\section{Exemplo de uso}
\label{sec:ex}

\begin{figure}[h!]
\begin{center}
\includegraphics[width=1.0\textwidth]{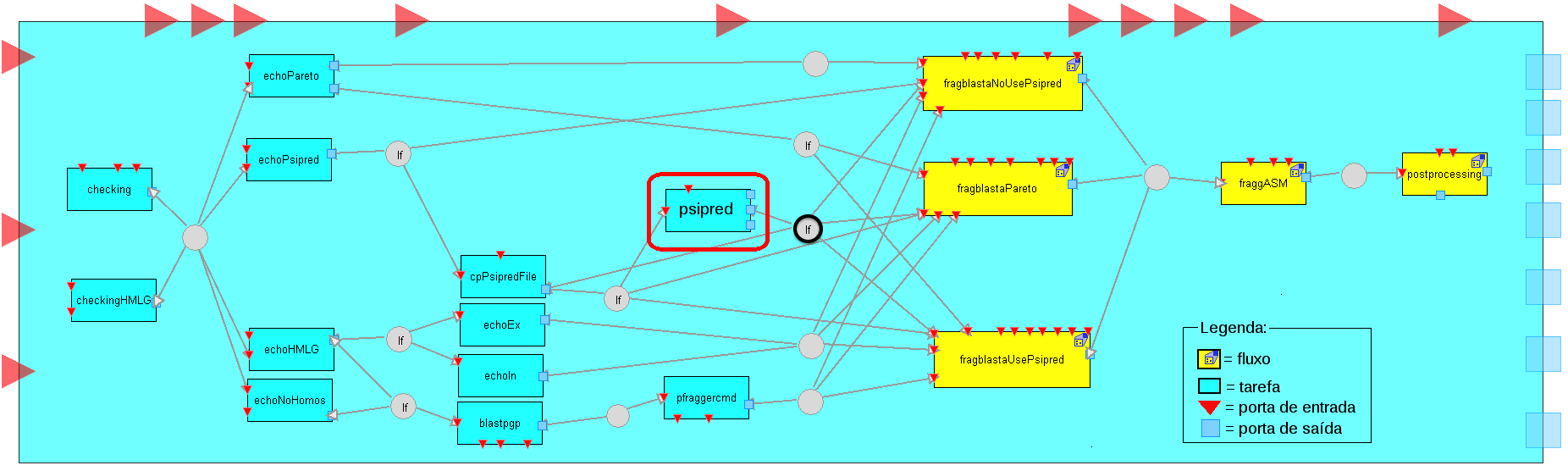}
\caption{Workflow Profrager}
\label{fig:profrager}
\end{center}
\end{figure}

O workflow ProFrager\footnote{\url{http://www.lncc.br/sinapad/Profrager/}} gera bibliotecas de fragmentos de proteínas. 
Este workflow foi escolhido para ilustrar a expressividade de OSC e para os testes do protótipo de SGWfC com suporte a OSC por ser o que demanda mais atributos não-funcionais dentre os workflows estudados neste trabalho.\footnote{Descrições completas de outros exemplos de uso de OSC, além do ProFrager, estão disponíveis em\\ \url{www.lncc.br/sinapad/OSC/examples.htm}.}
A representação gráfica desse modelo é ilustrada na Figura \ref{fig:profrager}. 

Por questões de espaço, somente a tarefa \textit{psipred} e o conector \textit{if3} -- realçados na Figura~\ref{fig:profrager} -- têm sua especificação em Acme apresentada neste trabalho (vide Figura~\ref{pfragger}).
A tarefa \textit{psipred} exemplifica ambos os atributos de qualidade existentes em OSC através de uma composição do tipo básico Executavel com os tipos Log, RedundanciaTemporal e OPM.
As propriedades \textit{num\_tentativas} e \textit{ignorar} pertencem ao tipo RedundanciaTemporal e a configuração criada permitiu que, durante os testes com o protótipo do SGWfC OSC, essa tarefa fosse ignorada nos casos em que apresentou falha após três tentativas de execução. 
Quando a tarefa \textit{psipred} é ignorada o conector \textit{if3}, o qual possui dois papéis de origem de dados, recebe o sinal de falha da tarefa no papel ligado a sua porta de saída e utiliza os dados vindos através do papel associado à tarefa \textit{cpPsipredFile} (vide Figura \ref{fig:profrager}). 

O tipo OPM, presente na descrição da tarefa e de suas portas, representa o modelo OPM para rastreamento de proveniência. 
A propriedade \textit{versao} configurada como \emph{orange} e \emph{black} informa que os dados de proveniência relacionados a esta tarefa e suas portas serão armazenados em ambas as versões que possuem estes nomes no grafo OPM. 

\begin{figure}[h!]
\begin{center}
\begin{lstlisting}
Component psipred : Log, Executavel, OPM, RedundanciaTemporal =	new RedundanciaTemporal, OPM, Executavel, Log extended with {
    //... portas

    Property ignorar = true;
    Property num_tentativas = 3;
    Property versao = {``orange'',``black''};
    //... outras propriedades da tarefa
}
Connector if3 : IfDados, CorrecaoPorPropagacaoCon, DistribuidoCon = 
	new IfDados, CorrecaoPorPropagacaoCon, DistribuidoCon extended with {
    Role psipred : Batch, Origem = new Origem, Batch extended with {
        //... propriedades
    }
    Role psipredFalso : Batch, Origem = new Origem, Batch extended with {
        //... propriedades
    }
    //... papeis de destino
    //... propriedades
}
\end{lstlisting}
\caption{Exemplo de tarefa e conector OSC compostos com atributos de qualidade}
\label{pfragger}
\end{center}
\end{figure}




\section{Trabalhos relacionados}
\label{sec:trab} 

A Tabela \ref{tab:tr_atr_nfunc} aponta o suporte a atributos não-funcionais em OSC e em cada um dos SGWfCs estudados \textbf{com relação à linguagem de descrição de workflows que adotam}. 
Como estes sistemas são extensíveis, alguns atributos não são suportados por padrão, mas possuem extensões que permitem a configuração dos atributos não-funcionais. 
Aqueles que não foram suportados, mas estão bem especificados na literatura, foram mencionados. 
Do contrário, foram anotados na tabela como sem suporte por padrão. 
Na maioria dos casos percebe-se que os trabalhos oferecem algum suporte a combinação de elementos funcionais do workflow com mecanismos de tratamento de atributos de qualidade, porém, as configurações destes mecanismos tipicamente  ou não são realizadas na descrição dos workflows ou são restritas a poucas opções.
Para o atributo de rastreabilidade, por exemplo, os SGWfCs comumente permitem a criação de anotações ou de forma geral para todo o workflow ou para cada tarefa. 
OSC é a única linguagem dentre as apresentadas que permite a inclusão dos tipos de rastreamento de proveniência por elemento do modelo, o que é uma característica vantajosa no que tange a configurabilidade do rastreamento, porém, para casos onde o usuário deseja realizar tal rastreamento para todo o fluxo, OSC apresenta um formato um pouco mais trabalhoso. 
Além disso, pelo que pôde ser levantado desses trabalhos, o suporte à configuração do mecanismo de tolerância a falhas nas interações é oferecido unicamente pelo OSC.
\begin{table}[h!]
 \tiny
 \centering
 \caption{Suporte a atributos não-funcionais nas linguagens utilizadas pelos SGWfCs.}

 \begin{tabular}{lp{3.0cm}lp{3.0cm}lp{3.0cm}|} 
 \hline

\textbf{SGWfC} & \textbf{Atributos não-funcionais} & \textbf{Configuração destes atributos no modelo}\\
 \hline
\textbf{Swift} & proveniência & Não é configurável na criação do modelo \citep{gadelhajr2011}.\\
 \citep{zhao2007} & tolerância a falhas & Os usuários podem configurar o tempo máximo para execução de cada tarefa.\\
 & paralelismo de tarefas & Os usuários podem configurar algumas opções, p. ex. o número de processos, pela configuração dinâmica de perfis.\\
 & varredura de parâmetros & Através do operador ``\textit{foreach}.''\\
 & MapReduce & Não foi encontrado suporte por padrão.\\
\\
\textbf{VisTrails} & proveniência & Os usuários podem adicionar notas às tarefas.\\
\citep{scheidegger2008} & tolerância a falhas & Não foi encontrado suporte.\\
 & paralelismo de tarefas & Não foi encontrado suporte por padrão.\\
 & varredura de parâmetros & Possui um modo específico em sua interface gráfica para este tipo de execução que\\
& & permite a configuração não só dos dados de entrada, como também dos resultados.\\
 & MapReduce & Parte deste mecanismo é suportado pelo módulo Map do VisTrails.\\
\\
\textbf{Taverna} & proveniência & Os usuários podem adicionar notas às tarefas e interações entre tarefas.\\
 \citep{missier2010} & tolerância a falhas & Usuários podem definir a quantidade de reexecuções das tarefas que apresentam erros.\\
 & paralelismo de tarefas & A configuração é permitida através de \textit{plug-ins} para execução remota de tarefas, e é distinta por \textit{plug-in}.\\
 & varredura de parâmetros & Se o usuário passa $N$ valores para portas que aceitam valores únicos, realiza a varredura de parâmetros por padrão.\\
 & MapReduce & O suporte está em desenvolvimento pelo projeto SCAPE. \\  
 & &\url{http://www.taverna.org.uk/introduction/related-projects/scape/}\\
\\
\textbf{Kepler} & proveniência & Os usuários podem configurar a proveniência somente para fluxos \citep{altintas2006}.\\
\citep{ludascher2006} & tolerância a falhas & Os usuários só podem configurar os mecanismos deste atributo nos fluxos como um todo \citep{mouallem2010}.\\
 & paralelismo de tarefas & Não foi encontrado suporte por padrão.\\
 & varredura de parâmetros & Configurável através do conjunto de atores e diretores desenvolvidos para a utilização \\
 & & do conjunto de ferramentas para grades computacionais chamada Nimrod \citep{abramson2009}.\\
 & MapReduce & Através da utilização do ator MapReduce\citep{wang2009}, desenvolvido para dar suporte \\
 & & ao Hadoop, o usuário pode compor seus fluxos utilizando este modelo.\\
\\
\textbf{OSC} & proveniência & Os usuários podem configurar os mecanismos de proveniência todos os elementos definindo níveis de granularidade.\\
 & tolerância a falhas & Os usuários podem configurar os mecanismos de tolerância a falhas para tarefas (incluindo fluxos) e conectores.\\
 & paralelismo de tarefas & Os usuários podem configurar suas tarefas para execução em ambientes de memória compartilhada e/ou distribuída.\\
 & varredura de parâmetros & Os usuários podem criar fluxos que dêem suporte a este atributo.\\
 & MapReduce & A implementação das tarefas internas aos fluxos deste tipo é que definem sua execução através deste mecanismo.\\
 \hline
 \end{tabular}
 \label{tab:tr_atr_nfunc}
 \end{table}

\vspace{-0.05in}
\section{Conclusão e Trabalhos Futuros}
\label{sec:conc}

A linguagem OSC faz uso da grande flexibilidade do sistema de tipos de Acme, o que a torna ao mesmo tempo facilmente extensível e propícia ao desenvolvimento para reuso, bem como permite ao cientista empregar esses tipos para compor a configuração de diferentes mecanismos de gerência dos atributos não-funcionais de interesse em um workflow.

A utilização neste artigo de powertypes da UML 2.0 para a exposição das formas de composição dos tipos ofertados pela linguagem OSC mascara algumas das dificuldades encontradas ao longo do desenvolvimento deste trabalho no que tange a definição das regras arquiteturais do estilo Acme que define OSC.
Em Acme, essas regras são definidas por meio de predicados em lógica de primeira ordem, que não têm expressividade suficiente para representar algumas restrições na combinação entre tipos da linguagem OSC.
A criação de funções de validação usando a biblioteca AcmeLib, que permite a manipulação programática de especificações Acme na linguagem Java, vendo sendo conduzida como parte deste trabalho para suprir essa limitação.



Como Acme foi concebida para o intercâmbio de descrições arquiteturais entre diferentes ferramentas de especificação arquitetural, ela se mostra particularmente adequada como base para a aplicação de técnicas de transformação de modelos.
No contexto deste artigo, pretende-se como trabalho futuro empregar essa característica de Acme para usar OSC também como linguagem para intercâmbio de especificações de workflows.


\vspace{-0.6in}
\small{
\bibliographystyle{sbc-br}
\bibliography{e-science13}
}

\end{document}